\documentclass[article]{appolb}
\usepackage{hyperref} 
\usepackage{multirow,graphicx}

\begin{document}
\title{{\WERS{Human Tissues Investigation\break Using PALS Technique}}%
\thanks{Presented at the $2^\mathrm{nd}$ Jagiellonian Symposium on Fundamental and Applied Subatomic Physics, Kraków, Poland, June 4--9, 2017.}}
\headtitle{Human Tissues Investigation Using PALS Technique}
\author{{B.}{Jasińska}$^\mathrm{a,}$\thanks{Corresponding author: \texttt{bozena.jasinska@umcs.pl}}, 
{B.}{Zgardzińska}$^\mathrm{a}$, 
{G.}{Chołubek}$^\mathrm{b}$, 
{M.}{Gorgol}$^\mathrm{a}$\\ 
{K.}{Wiktor}$^\mathrm{c}$, 
{K.}{Wysogląd}$^\mathrm{a}$, 
{P.}{Białas}$^\mathrm{d}$, 
{C.}{Curceanu}$^\mathrm{e}$\\ 
{E.}{Czerwiński}$^\mathrm{d}$, 
{K.}{Dulski}$^\mathrm{d}$, 
{A.}{Gajos}$^\mathrm{d}$, 
{B.}{Głowacz}$^\mathrm{d}$\\ 
{B.}{Hiesmayr}$^\mathrm{f}$, 
{B.}{Jodłowska-Jędrych}$^\mathrm{g}$, 
{D.}{Kamińska}$^\mathrm{d}$\\ 
{G.}{Korcyl}$^\mathrm{d}$, 
{P.}{Kowalski}$^\mathrm{d}$, 
{T.}{Kozik}$^\mathrm{d}$, 
{N.}{Krawczyk}$^\mathrm{d}$\\ 
{W.}{Krzemień}$^\mathrm{h}$, 
{E.}{Kubicz}$^\mathrm{d}$, 
{M.}{Mohammed}$^\mathrm{d,i}$\\ 
{M.}{Pawlik-Niedźwiecka}$^\mathrm{d}$, 
{S.}{Niedźwiecki}$^\mathrm{d}$, 
{M.}{Pałka}$^\mathrm{d}$\\ 
{L.}{Raczyński}$^\mathrm{j}$, 
{Z.}{Rudy}$^\mathrm{d}$, 
{N.G.}{Sharma}$^\mathrm{d}$, 
{S.}{Sharma}$^\mathrm{d}$, 
{R.}{Shopa}$^\mathrm{j}$\\ 
{M.}{Silarski}$^\mathrm{d}$, 
{M.}{Skurzok}$^\mathrm{d}$, 
{A.}{Wieczorek}$^\mathrm{d}$, 
{H.}{Wiktor}$^\mathrm{k}$\\ 
{W.}{Wiślicki}$^\mathrm{j}$, 
{M.}{Zieliński}$^\mathrm{d}$, 
{P.}{Moskal}$^\mathrm{d}$
\address{$^\mathrm{a}$Institute of Physics, Maria Curie-Skłodowska University, Lublin, Poland\\
$^\mathrm{b}$Diagnostic Techniques Unit, Faculty of Nursing and Health Sciences
Medical University of Lublin, Poland\\
$^\mathrm{c}$Dept. of Obstetrics, Gynaecology and Obstetrics\,---\,Gynaecological Nursing\\
Faculty of Nursing and Health Sciences, Medical University of Lublin, Poland\\
$^\mathrm{d}$Faculty of Physics, Astronomy and Applied Computer Science\\
Jagiellonian University, Kraków, Poland\\
$^\mathrm{e}$INFN, Laboratori Nazionali di Frascati, Frascati, Italy\\
$^\mathrm{f}$Faculty of Physics, University of Vienna, Vienna, Austria\\
$^\mathrm{g}$Chair and Department of Histology and Embryology\\ 
with Experimental Cytology, Medical University of Lublin, Poland\\
$^\mathrm{h}$High Energy Physics Division, National Centre for Nuclear Research\\
Otwock-Świerk, Poland\\
$^\mathrm{i}$Department of Physics, College of Education for Pure Sciences\\
University of Mosul, Mosul, Iraq\\
$^\mathrm{j}$Department of Complex Systems, National Centre for Nuclear Research\\
Otwock-Świerk, Poland\\
$^\mathrm{k}$Chair and Department of Gynaecology and Gynaecological Endocrinology\\
Faculty of Nursing and Health Sciences, Medical University of Lublin, Poland\\}}
\headauthor{B. Jasińska et al.}
\maketitle
 
 \newpage
\begin{abstract}
Samples of uterine leiomyomatis and normal tissues taken from patients after surgery were investigated using the Positron Annihilation Lifetime Spectroscopy (PALS). 
Significant differences in all PALS parameters  between normal and diseased tissues were observed. For all studied patients, it was found that the values of the free annihilation and 
ortho-positronium lifetime are larger for the tumorous tissues than for the healthy ones. For most of the patients, the intensity of the free annihilation and ortho-positronium 
annihilation was smaller for the tumorous than for the healthy tissues.
For the first time, in this kind of studies, the $3\gamma$ fraction of positron annihilation was determined to describe changes in the tissue porosity during morphologic alteration.
 
\end{abstract}


\section{Introduction}

For the last few decades, a positron has become a valuable tool in material investigations. Experimental technique based on the positron behaviour in the medium, Positron Annihilation Lifetime Spectroscopy (PALS),  
is commonly used in investigations of various kind of materials: from metals to complex mesoporous materials~\cite{1, 2, 3}. The thermalized positron produced initially during $\beta$ decay can directly annihilate with one of the electrons or create positronium (Ps) \ie a bound state of a positron and an electron. Positronium can exist in two different states: para-positronium (p-Ps) and ortho-positronium (o-Ps), both substates: p-Ps and o-Ps are not stable and annihilate in the vacuum with a mean lifetime values: 125 ps and 142 ns, respectively. In the medium however, the o-Ps lifetime value can be shortened even below 1\,ns due to the possibility of the positronium trapping in the regions of lower electron density, the so-called free volume or void.  In these voids,  o-Ps can annihilate not only by intrinsic decay but with one of the electrons from its surroundings by  
pick-off process~\cite{4}. Shortening of the o-Ps lifetime value is determined by the local electron density, which is correlated with the size of the void where the positronium is trapped~\cite{5,6}. The equation presented below describes the dependence of the o-Ps lifetime value $\tau_\mathrm{pick-off}$ on the void radius $R$ in which positronium is trapped
\begin{equation}
\tau_\mathrm{pick-off} = \frac{1}{\lambda_b}\left(1 - \frac{R}{R + {\mit\Delta}} + \frac{1}{2\pi} \sin \left(\frac{2\pi R}{R + {\mit\Delta}}\right) \right)^{-1}\,, \qquad \lambda_b = 2~\mathrm{ns}^{-1}\,,
\label{eq1}
\end{equation}
where $R$ is the void radius, $\lambda_b$ --- averaged Ps decay constant, ${\mit\Delta}$ --- empirical parameter reflecting  Ps interaction with the surroundings. The formulas relating o-Ps lifetime with the void size~\cite{7,8,9,10,11,12,13} allow to use the o-Ps lifetime value to determine  the void sizes in the range  from 0.2 to about 100 nm.\break
\newpage
\noindent The o-Ps lifetime in the living organism should be  connected to the morphology of the cells and may be used as an  indicator of the stage of the development of metabolic disorders~\cite{27}.

The other parameter that can be applied to study the porosity of the material/tissues modification  is the fraction $f_{\mathrm{o-Ps}-3\gamma}$ of o-Ps atoms annihilating with the emission of $3\gamma$ quanta~\cite{28,39}. Taking into account two processes leading to o-Ps decay: intrinsic decay and pick-off process, the $3\gamma$ fraction can be described as $f_{\mathrm{o-Ps}-3\gamma}= \tau_\mathrm{o-Ps}/\tau_T$, where $\tau_T$ denotes the o-Ps lifetime value in the vacuum, equal to 142 ns. In the case when only one kind of  free volumes/pores exists in the material, the $3\gamma$ fraction in the whole spectrum can be effectively described by the dependence
\begin{equation}
f_{3\gamma} = \frac{\left(1 -\frac{4}{3}P_\mathrm{o-Ps}\right)}{372} + \frac{\tau_\mathrm{o-Ps}}{\tau_T}P_\mathrm{o-Ps}\,,
\label{eq2}
\end{equation}
where, $P_\mathrm{o-Ps}$ denotes the ortho-positronium formation probability depending on the molecular structure of the investigated object. It was successfully applied to investigate 
high-porosity materials~\cite{14,15,16,17}. In this article, we describe an application of o-Ps lifetime and intensity as well as the first application of the $3\gamma$ fraction for investigation of the  diseased tissues of human body. 

The PALS was successfully applied in studies of many classes of materials but only in a very limited number of papers concerning living biological systems. Some applications were described by Jean and Ache in 1977~\cite{18}. They focused on studies of healthy and abnormal skin samples~\cite{19,20} and reported that the $S$ parameter from the Doppler broadening of annihilation line is correlated with a broadly defined level of skin damage. For the last few years, biological systems have aroused the interest of annihilation techniques again~\cite{21}. The precise investigations of human tissues seems to be a very complex problem because of the presence of biofluids where positronium  can also undergo annihilation. However, hydrated solid materials were successfully studied using PALS. In the paper by Hugenschmidt \etal \cite{22}, some experiments concerning the behaviour of the free volume on water loading, drying, and uniaxial pressure on glucose--gelatine compounds were performed. The dynamics of the water sorption by hydrophilic lyophilised yeast cells was studied~\cite{23}. The precise understanding of various processes in biological systems is very important, as they can be applied to increase the possibilities of human body diagnosis using the Positron Emission Tomography (PET). One of the innovative  scanner (referred to as J-PET) has been constructed at the Jagiellonian University,  Kraków, Poland~\cite{24}. One of the advantages of J-PET is a possibility of multi-photon imaging~\cite{25, 26} which enables  the diagnosis based on positron and positronium lifetime~\cite{27} as well as based on the ratio of $3\gamma$ and $2\gamma$ annihilation rates~\cite{28}.

In this paper, the human normal and diseased tissues are investigated using PALS in order to compare modifications in the tissue structure during tumour progression.

\section{Experimental}

\subsection{Materials}

Uterine leiomyomas are the most common benign uterine tumours in women. Histologically, they consist mainly of smooth muscle cells and contain different amounts of fibrous tissue. 
Literature data have shown that more than 30\% of women have myomas~\cite{29}. 

In our study, patients with myomas were qualified to a surgery on the basis of symptoms, history, clinical examination and ultrasonography. Each patient underwent hysterectomy with or without adnexa. The study was approved by the Bioethics Committee of the Medical University of Lublin. Each patient agreed to a clinical trial. Samples were taken directly from the organ, just after removal of the uterus from the surgical field by one of the two experienced researchers in the operating room. Tissue fragments were taken from the sites rated as macroscopically altered as well as from the normal tissues and placed directly into a transporting chamber. In any case, the time of delivery of the tissue to testing, in the same stable temperature ($23\pm 1^\circ$C) was less than 1 hour. Each of the examined tissues was subsequently subjected to a histopathological examination (Table I).

\begin{table}[htb]
{\small
\rightline{TABLE I}
\vspace{3mm}
\centerline{Comparison of the tested samples.}}

\vspace{3mm}
\centerline{%
\includegraphics[width=12.5 cm]{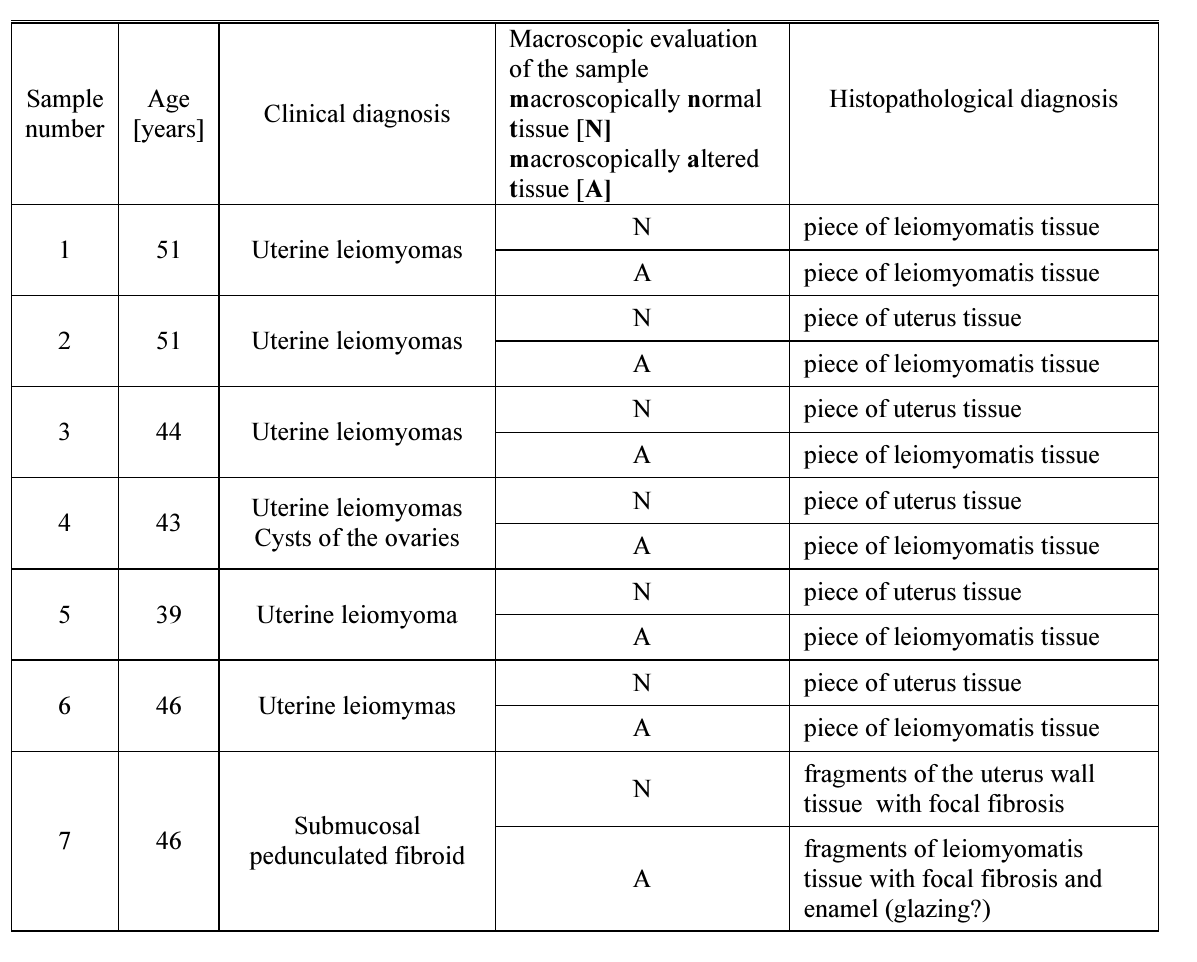}}
\end{table}

\subsection{Methods}

\begin{figure}[htb]
\centerline{%
\includegraphics[width=8cm]{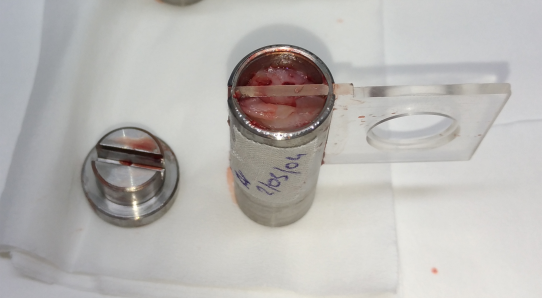}}
\caption{Open chamber with tissue sample. The $^{22}$Na source was placed in the hole in the plexiglass plate and then inserted into the sample area (in this case: left-hand side).}
\label{fig1}
\end{figure}

To perform PALS measurements and compare the results for samples with tumor and the reference one, two  slices of each sample, separated by a plexiglass partition element, were placed inside the steel chamber presented in  Fig.~\ref{fig1}. 
The total volume of the sample was about 1.5 cm$^3$. The $^{22}$Na positron source with the activity of 700 kBq was placed inside the partition and inserted between the slices of the sample directly before the measurement. The steel seal assured that the sample was closely attached to the source. The sample chamber was placed between two scintillation heads, equipped with BaF$_2$ scintillators, used  as START and STOP counters. The detectors collected the $\gamma$ quanta corresponding to positron creation inside the source and positron/positronium annihilation inside the sample, respectively. The electrical impulses from the heads were processed with the use of standard Fast--Slow delayed coincidence spectrometer~\cite{1,3}. The time intervals between START and STOP events were measured with the use of time-to-amplitude converter with the time range of 100 ns. The resolution curve of the spectrometer was approximated by a single Gaussian with the full width at half maximum (FWHM) of 230 ps. Such parameters allowed to perform precise measurements of positron/positronium lifetimes from about 100 ps, up to few ns. 

In order to exclude the influence of sample aging on PALS results, the reference sample, and the one with the tumor were measured at the same time, with the use of two PALS spectrometers of almost identical parameters and the positron sources of similar activities. After 1.5 hours of measurements, the samples were swapped, and then, the data from two spectrometers were compared. The analysis of the results showed that:
\begin{itemize}
\item[---] PALS parameters do not depend on the spectrometer;
\item[---] In the interval of a few hours, no influence of sample aging on PALS results was noticed.
\end{itemize}

The lifetime spectra were analyzed with the use of \textsf{LT 9.2} program~\cite{30} and three discrete components ascribed to annihilation of para-positronium (p-Ps, $\tau_1$), 
unbound positrons ($e^+$, $\tau_2$) and ortho-positronium (o-Ps, $\tau_3$) were found, respectively, for each sample.

\section{Results and discussion}

\begin{figure}[htb]
\centerline{%
\includegraphics[width=6.2cm]{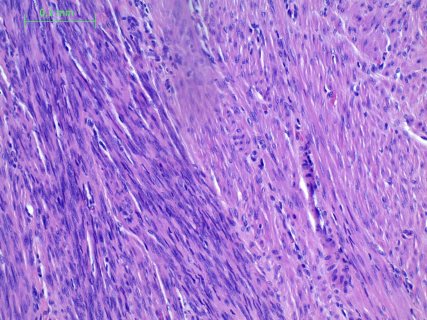}
\includegraphics[width=6.2cm]{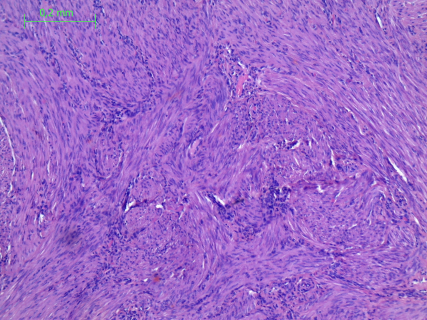}}
\caption{Linear muscle cells of a normal uterus (left) and chaotic fibroid cells (right) --- both images  from the tested tissues. 
Magnification is equal to 100. Figure length: 0.62 mm and 1.2 mm, respectively.}
\label{fig2}
\end{figure}

For each patient, a pair of samples was investigated: one normal (N) and one tumor-altered (A).  Exemplary microscopic images for one of the pair of tissues are presented in Fig.~\ref{fig2}. 
The differences in organization and density between the healthy and tumorous tissues are well-visible  in the micrometer scale. 

\begin{figure}[b]\vspace{-3mm}
\centerline{%
\includegraphics[width=12.5cm]{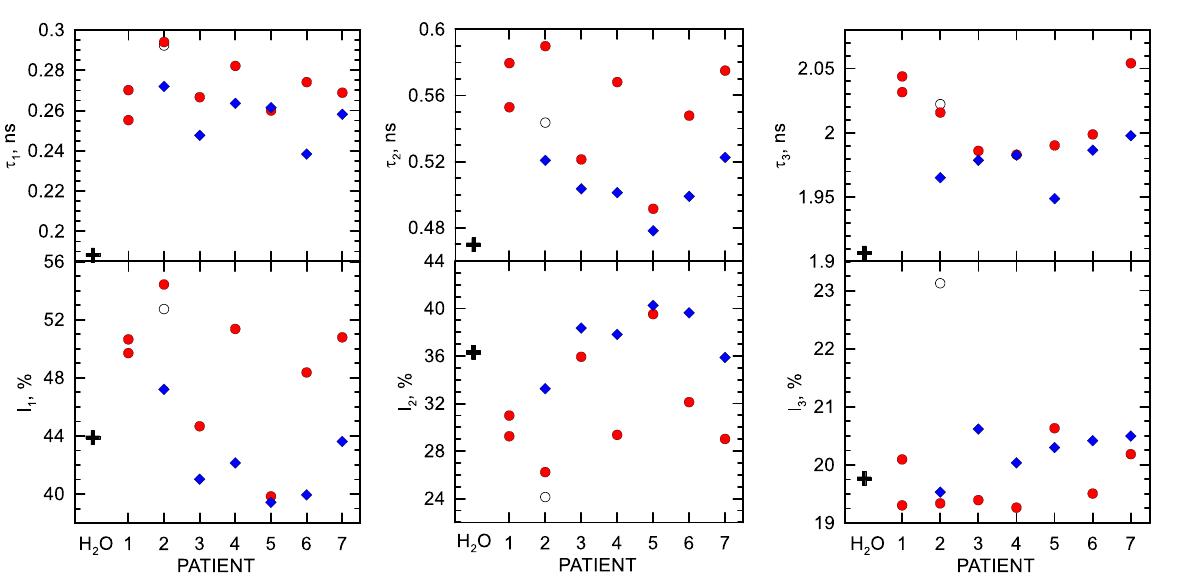}
\put(-295,-10){\scriptsize{(a) \hspace{11.8em} (b) \hspace{12.5em} (c)}}}
\caption{(Colour on-line) The lifetime and intensity values of PAL spectra components for normal (blue diamonds) and altered (red circles) uterine tissue. From left to right: p-Ps, free annihilation, o-Ps. In addition: uterine leiomyomas tissue after 16~hours storage in formalin (empty circles) and deionized water at room temperature (crosses). For the first patient, both samples were classified as altered by means of the microscopic images.}
\label{fig3}
\end{figure}

\newpage
The results of PAL spectra analysis for normal (diamond) and altered (circle) tissues are shown  in Fig.~\ref{fig3}. 
For comparison,  the reference points for water (crosses), which is the main constituent of biofluids in living organisms is added. Water was extensively tested by the PALS technique~\cite{31,32,33,34,35,36,37}, but for the purpose of this study, additional measurement for deionized water was carried out under the same measurement conditions as for human tissues.  During the \textsf{LT} analyses, all three components were assumed as free. For human tissues, there are no justified reasons to fix the ratio $I_1:I_3 = 1:3$ or to fix the p-Ps value at 125 ps (it was found that the use of such procedures significantly modifies the results). This kind of spectra analysis leads to some parameter misrepresentation, especially the first component intensity is unviable high, however, it is a justified method in this stage of investigation. Moreover, the main purpose of this paper is a comparison between normal and altered tissues taken from the same  patient and to study the  
patient-to-patient variations. From the results presented in  Fig.~\ref{fig3}, one can conclude that there are significant differences of results obtained for normal (and also altered) tissues between different female patients. Therefore, it is not possible to determine the state of health of the tissue from a single PAL spectrum measurement, and it is necessary to correlate two measurements performed for normal and altered tissues from the same patient. The differences in the absolute values of the PAL spectra parameters are small but detectable. While correlating the respective parameters values obtained for normal and altered tissues from the same patient, a clear trend can be observed: the lifetime values of all three components in altered tissue are higher than those measured in normal tissue from the same patient. Similar relation was observed previously in the literature~\cite{19,38}.

The PALS parameters measured in tissues are similar, but not equal, to those known for water. While the intensity of the components in these two media are very close, 
differences are noticeable in the values of all three lifetimes --- in tissues all lifetimes are longer. 
The greatest discrepancies were found for the p-Ps lifetime: ($0.188 \pm 0.006$) ns and ($0.272 + 0.008$) ns for water and normal tissue sample, respectively. 

The comparison of PALS parameters for fresh tissue samples and the ones conserved in formalin was also performed. The altered tissue from the second patient after the standard measurement was placed for 16 h in formalin at $23^\circ$C and then measured again with the use of PALS technique. It was found that the formalin storage significantly influenced the studied parameters, especially the o-Ps intensity (an increase from the initial value 19.5\% up to 23\% was observed) and the average lifetime of the second component related to the annihilation of free positrons. Based on the measurement presented above, it was concluded that samples intended for PALS measurements should not be modified in advance to obtain characteristics for the  possible future \textit{in vivo} imaging as results should reflect composition of living organisms. 

\begin{figure}[htb]
\centerline{%
\includegraphics[width=11cm]{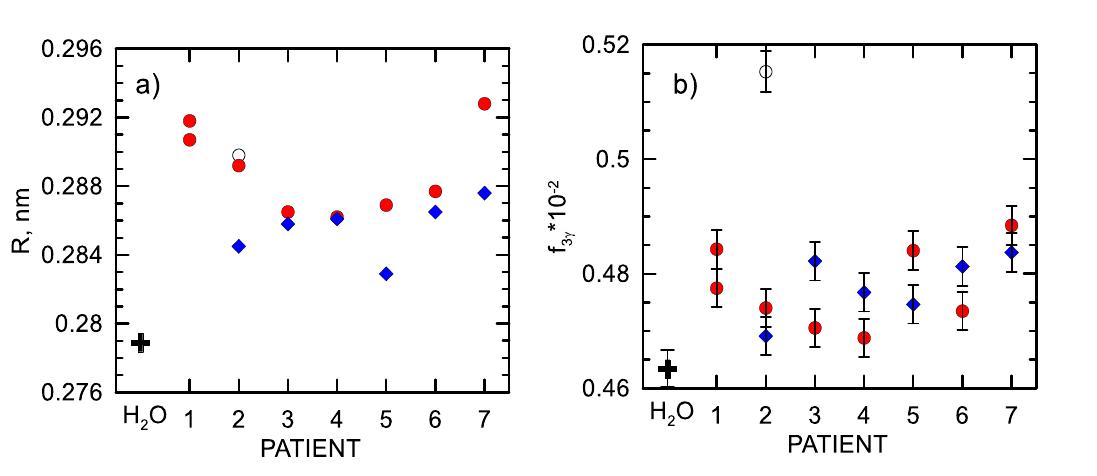}}
\caption{(Colour on-line) The void radius (left), and $3\gamma$ fraction (right),  as estimated from PALS, determined for normal (blue diamonds) and altered (red circles) tissues.}
\label{fig4}
\vspace{-3mm}\end{figure}

From data presented in Fig.~\ref{fig3}\,(a) the void radius (assuming spherical shape of void) and (b)  the fraction of $3\gamma$ positron annihilation were calculated according to equations (\ref{eq1}) 
and (\ref{eq2}), respectively (Fig.~\ref{fig4}). The difference in both parameters between the normal and diseased tissues are small and varies from patient to patient: in the altered tissues, voids are sparingly larger, while $3\gamma$ fraction change non-monotonically. The $3\gamma$ fraction depends on both: lifetime and intensity of o-Ps component which reflect the void size and the concentration, respectively. In effect, it depends on the porosity structure of the investigated material. In the case of our measurements, it may suggest that the observed difference in PALS parameters between normal and altered tissues reflects the degree or the kind of tissue pathogenic modification. This conclusion seems to be confirmed by measurements performed  on the first sample: the material microscopically considered to be a normal tissue after the histological examination was qualified as diseased (Table I); therefore, PALS results reflect the degree of tissue deformation in one patient (involved with the experimental uncertainty).

\newpage

\section{Conclusions}

The preliminary results of PALS measurements for pairs of samples: normal and diseased, taken from patients' uterus just after surgery and not solidified show small but meaningful differences inside the pair and between the patients.  For all studied patients, the free annihilation and ortho-positronium lifetime values were found to be larger for the diseased tissues with respect to the lifetime of the normal ones. At the same time, the intensity of both components, generally, was found to be smaller for altered tissues than for the normal ones.  Our results that provide the information on the molecular scale of the tissues structure justify the expectation  that PALS measurements of positron and positronium parameters could be useful in the cancer/tumour diagnostics. They constitute the first step on a way to establish correlations between  positronium properties in the tissue and the staging of tumours for the elaboration of the \textit{in vivo} morphometric 
imaging proposed in~\cite{27, 28}.
 
However, to unequivocally conclude the existence of such correlations, a large number of patients and different kinds of tumours/cancers have to be analyzed.

\vspace{7mm}
We acknowledge the support by the National Science Centre, Poland (NCN) through the grant No. 2016/21/B/ST2/01222, B.C.H. --- the Austrian Science Fund (FWF-P26783).

\newpage
\flushleft

\end{document}